\documentclass[12pt]{article}
\usepackage{graphicx,epsfig,cite}
\usepackage{epsfig,amssymb,amsmath}
\usepackage{graphics,subfigure}
\begin{document}

\title{{\normalsize\begin{flushright}
  IPPP/06/25\\
  DCPT/06/50\\
  CERN-PH-TH/2006-070\\
  hep-ph/0607104
\end{flushright}}
\vskip 1cm
Combined LHC/ILC analysis of a SUSY scenario with heavy sfermions}
%
\author{\normalsize
{\bf
K.~Desch$^{a}$,
J.~Kalinowski$^{b}$,
G.~Moortgat-Pick$^{c}$,
K.~Rolbiecki$^{b}$, W.J.~Stirling$^{d}$
}\\[3mm]
$^a${\normalsize\it Physikalisches Institut, Universit\"at Bonn,
  D-53115   Bonn, Germany}\\
$^b${\normalsize\it Instytut Fizyki Teoretycznej, Uniwersytet Warszawski,
  PL-00681   Warsaw, Poland}\\
$^c${\normalsize\it TH Division, Physics Department, CERN, CH-1211 Geneva 23,
  Switzerland}\\
$^d${\normalsize\it IPPP, University of Durham, Durham DH1 3LE, UK}
}

\maketitle

\begin{abstract}
  We discuss the potential of
analyses at the Large Hadron Collider
  and the planned International Linear Collider to explore low-energy
  supersymmetry in a difficult region of the parameter space characterized by
  masses of the scalar SUSY particles around 2~TeV. Precision
  analyses of cross sections for light chargino production and
  forward--backward
  asymmetries of decay leptons and hadrons at the { first stage of the ILC
 with $\sqrt{s}=500$~GeV},
  together with mass information on ${\tilde{\chi}^0_2}$ and squarks from
 the LHC,
allow us to determine the underlying fundamental gaugino/higgsino MSSM
  parameters and to constrain the masses of the heavy, kinematically
  inaccessible sparticles.
No assumptions on a specific SUSY-breaking
  mechanism
  are imposed.  For this analysis the complete spin correlations between
  production and decay processes { must be} taken into account.
\end{abstract}

\section{Introduction}

Supersymmetry (SUSY) is one of the best-motivated candidates for physics
beyond the Standard Model (SM). Low-energy SUSY is well-motivated since it
stabilizes the electroweak scale, provides quantitatively accurate
unification of gauge couplings as well as a promising cold-dark-matter
candidate.
Moreover, { electroweak precision data \cite{ewwg}
and cosmology bounds \cite{wmap}
seem to indicate  \cite{Ellis:2006ix}
that at least some of the electroweakly interacting
SUSY particles should be rather light and accessible at future colliders}.
However, since the mechanism of SUSY breaking is unknown,
supersymmetric extensions of the Standard Model contain a large number of
unknown parameters, e.g.\ 105 in the Minimal Supersymmetric Standard Model
(MSSM).
Specific assumptions on the SUSY-breaking mechanism, in particular
about the unification of parameters at the grand-unification (GUT) scale,
considerably reduce the number of free parameters, e.g.\ in
the constrained MSSM,
{ often referred to as mSUGRA}, where we end up
with only four new parameters (and one sign) specified at the
unification scale. Experiments at future accelerators, the Large Hadron
Collider (LHC) and
the International Linear Collider (ILC), will have,
however, not only to discover SUSY but also to
determine precisely the underlying  SUSY-breaking scenario with as
few theoretical prejudices as possible.

{ Particularly challenging are scenarios where the scalar SUSY
particle sector is very heavy as required, for instance, in focus-point
scenarios (FP)~\cite{focuspoint} where the  gaugino masses are kept relatively
small while  squarks and
sleptons might be too heavy for a direct observation at the ILC.  It is
therefore particularly interesting to verify whether the interplay
of a combined LHC/ILC analysis~\cite{Weiglein:2004hn} could shed light
on models with heavy sfermions}.

Many methods have been worked out to derive the SUSY parameters
at collider
experiments~\cite{Tsukamoto:1993gt,Feng:1995zd,Bechtle:2004pc,
  Lafaye:2004cn,Arkani-Hamed:2005px}.
In
\cite{Baer:1995tb,Choi:1998ut,Choi:2000ta,Kneur:1999nx,
  Choi:2001ww,Desch:2003vw} the
chargino and neutralino sectors have been exploited at the ILC to determine
the MSSM
parameters. However, in most cases only the production processes
have been studied and, furthermore, it has been assumed that the
masses of the virtual scalar particles are already known.
In the case of heavy scalars such assumptions cannot be applied and
further observables have to be used to determine the underlying
parameters.
Studies have been made to exploit the whole production-and-decay process, and
angular and energy distributions of
the decay products in chargino as well as
{ neutralino channels have been studied in
\cite{Moortgat-Pick:1998sk,Moortgat-Pick:1997ny,Barr:2004ze,Wang:2006hk}.}
Since such
observables depend strongly on the polarization of the decaying
particle, the spin correlations between production and decay
can have a large influence and have
to be taken into account.
Exploiting such spin effects, it has been shown
in~\cite{Moortgat-Pick:1999ck} that, once the chargino parameters are known,
useful indirect
bounds for the mass of the heavy virtual particles could be derived from
forward--backward asymmetries of the final lepton $A_{\rm FB}(\ell)$.

In this paper  we discuss a FP-inspired scenario characterized by a
$\sim$ 2 TeV scalar particles sector. In addition, the neutralino
sector turns out to have very low production cross sections in
$e^+e^-$ collisions, so that it might not be fully exploitable. Only
the chargino  pair production process has high rates at the ILC and
all information obtainable from this sector has to be used.  In
order to assess the possibility of unraveling such a challenging new
physics scenario, our analysis is performed entirely at the EW scale,
without any reference to the underlying SUSY-breaking mechanism. We
measure at the LHC and at the ILC with $\sqrt{s}=500$~GeV the
masses, cross sections and spin-dependent forward--backward
asymmetries and analyze the potential of a multiparameter fit to
determine the underlying parameters.

The paper is organized as follows. In the next section we first
present the studied process, chargino production with leptonic and
hadronic decays. We briefly introduce the spin formalism, which is
needed for the evaluation of spin-dependent observables. In  section
3 the FP-inspired scenario is defined and the expected experimental
results at the LHC and the ILC are discussed. In section 4 we
perform our numerical analysis and determine the SUSY parameters
based on the experimental input. An attempt at testing the $SU(2)$
symmetry relation for the selectron and sneutrino masses using the
available information on the squark masses from the LHC and the
forward--backward asymmetry measured at the ILC in hadronic decay
modes is also discussed. Section 5 summarizes the results.

\section{Strategy overview}
\subsection{Chargino and neutralino sector}
\begin{figure}
\hspace{1cm}
\begin{minipage}[t]{3.5cm}
\begin{center}
{\setlength{\unitlength}{1cm}
\begin{picture}(2.5,2.5)
\put(-.6,-1.1){\includegraphics{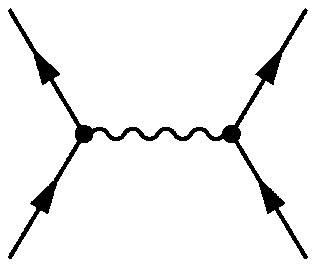}}
\put(-.6,-1.5){$e^{-}$}
\put(2.7,-1.5){$\tilde{\chi}^{-}_1$}
\put(-.6,1.5){$e^{+}$}
\put(2.7,1.5){$\tilde{\chi}^{+}_1$}
\put(1,.4){$\gamma$}
\end{picture}}
\end{center}
\end{minipage}
\hspace{2cm}
\vspace{.8cm}
\begin{minipage}[t]{3.5cm}
\begin{center}
{\setlength{\unitlength}{1cm}
\begin{picture}(2.5,2.5)
\put(-1.2,-1.1){\includegraphics{prog.ps}}
\put(-1.1,-1.5){$e^{-}$}
\put(2,-1.5){$\tilde{\chi}^{-}_1$}
\put(-1.1,1.5){$e^{+}$}
\put(2,1.5){$\tilde{\chi}^{+}_1$}
\put(.4,.4){$Z^0$}
\end{picture}}
\end{center}
\end{minipage}
\hspace{2cm}
\vspace{.8cm}
\begin{minipage}[t]{3.5cm}
\begin{center}
{\setlength{\unitlength}{1cm}
\begin{picture}(2.5,2)
\put(-1.2,-1.3){\includegraphics{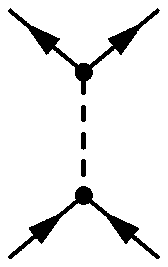}}
\put(-1.6,-1.5){$e^{-}$}
\put(-1.6,1.5){$e^{+}$}
\put(.4,-1.5){$\tilde{\chi}^{-}_1$}
\put(.4,1.5){$\tilde{\chi}^{+}_1$}
\put(-.9,0){$\tilde{\nu}$}
 \end{picture}}
\end{center}
\end{minipage}

\vspace{.5cm}

\begin{minipage}[t]{3.5cm}
\begin{center}
{\setlength{\unitlength}{1cm}
\begin{picture}(5,2.5)
\put(+.8,-.9){\includegraphics{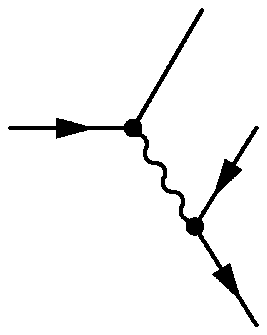}}
\put(3.9,-1.3){$\nu$, $q_u$}
\put(.8,1.1){$\tilde{\chi}^{+}_1$}
\put(3.7,1){$\ell^{+}$, $\bar{q}_d$}
\put(3.3,2.2){$\tilde{\chi}^0_1$}
\put(1.9,.1){$W^{+}$}
\end{picture}}
\end{center}
\end{minipage}
\hspace{2cm}
\vspace{.8cm}
\begin{minipage}[t]{3.5cm}
\begin{center}
{\setlength{\unitlength}{1cm}
\begin{picture}(5,4.5)
\put(+.3,+1){\includegraphics{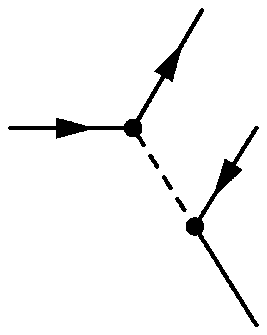}}
\put(2.5,4.2){$\nu$, $q_u $} \put(2.9,3){$\ell^{+}$, $\bar{q}_d$}
\put(.1,3.1){$\tilde{\chi}^{+}_1$} \put(3.1,.8){$\tilde{\chi}^0_1$}
\put(1.,2){$\tilde{\ell}_{{\rm L}}^*$, $\tilde{q}_{d{\rm L}}^*$}
\end{picture}}
\end{center}
\end{minipage}
\begin{minipage}[t]{3.5cm}
\begin{center}
{\setlength{\unitlength}{1cm}
\begin{picture}(5,4)
\put(1.5,.5){\includegraphics{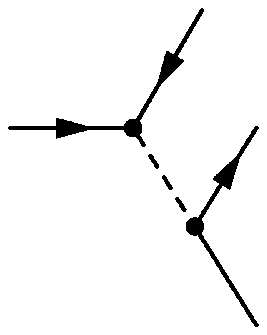}}
\put(4.2,2.5){$\nu$, $q_u$}
\put(3.7,3.6){$\ell^{+}$, $\bar{q}_d$}
\put(4.3,.4){$\tilde{\chi}^0_1$}
\put(1.3,2.6){$\tilde{\chi}^{+}_1$}
\put(2.3,1.5){$\tilde{\nu}$, $\tilde{q}_{u{\rm L}}$}
\end{picture}}
\end{center}
\end{minipage}
\vspace*{-1cm} \caption{\it Feynman diagrams for production and
for leptonic and hadronic decays of
  charginos. \label{fig_feyn}}
\end{figure}

We study chargino production
\begin{equation}
e^{-}+e^{+} \to \tilde{\chi}^{+}_1+\tilde{\chi}^{-}_1,\label{eq:p1}
\end{equation}
with subsequent leptonic and hadronic decays
\begin{eqnarray}
\tilde{\chi}^{+}_1 &\to&
\tilde{\chi}^0_1+\ell^{+}+\nu \quad\mbox{and}\quad
\tilde{\chi}^0_1+\bar{q}_d+q_u,\label{eq:p2}\\
\tilde{\chi}^{-}_1 &\to&
\tilde{\chi}^0_1+\ell^{-}+\bar{\nu} \quad\mbox{and}\quad
\tilde{\chi}^0_1+q_d+\bar{q}_u,\label{eq:p3}
\end{eqnarray}
where $\ell=e,\mu$, $q_u=u,c$, $q_d=d,s$. The corresponding Feynman
diagrams are shown in Fig.~\ref{fig_feyn}. The production process
contains contributions from $\gamma$- and $Z^0$-exchange in the
$s$-channel and from $\tilde{\nu}$-exchange in the $t$-channel. The
decay processes are mediated by $W^{\pm}$, $\tilde{\ell}_{\rm L}$,
$\tilde{\nu}$ or  by $\tilde{q}_{d{\rm L}}$, $\tilde{q}_{u {\rm L}}$
exchange; contributions from Higgs boson exchanges to the production and decay
are
negligibly small for the first and second generation fermions. { For
notations, couplings and conventions see, for instance~\cite{Haber:1984rc}.}

The neutralino
mass eigenstates are defined as $\tilde{\chi}^0_i=N_{ij}
\tilde{\lambda}_j$, where $N_{ij}$ are the elements of the unitary
$4\times 4$ matrix which diagonalizes the neutral gaugino--higgsino
mass matrix in the basis $\tilde{\lambda}=(\tilde{B}^0,
\tilde{W}^0, \tilde{H}^0_1, \tilde{H}^0_2)$:
\begin{eqnarray}
M_N &=&
\left(\begin{array}{cccc}
M_1 & 0
& -m_Z\cos\beta\sin\theta_W & m_Z\sin\beta\sin\theta_W \\
0& M_2 & m_Z\cos\beta\cos\theta_W & -m_Z\sin\beta\cos\theta_W\\
-m_Z\cos\beta\sin\theta_W & m_Z\cos\beta\cos\theta_W & 0& -\mu\\
m_Z\sin\beta\sin\theta_W & -m_Z\sin\beta\cos\theta_W & -\mu & 0
\end{array}\right),\nonumber\\
& &\label{1E}
\end{eqnarray}
where $m_Z$ denotes the mass of the $Z^0$ boson, $M_1$, $M_2$ are the
$U(1)$, $SU(2)$ gaugino mass parameters, $\mu$ is the Higgs mass
parameter and
$\tan\beta={v_2}/{v_1}$, where $v_{1,2}$ are the vacuum
expectation values of the two neutral Higgs fields.
The chargino mass eigenstates
$\tilde{\chi}_i^{+}={\chi_i^{+} \choose \bar{\chi}_i^{-}}$ are
defined by $\chi^{+}_i=V_{i1}w^{+}+V_{i2} h^{+}$ and
$\chi_j^{-}=U_{j1}w^{-}+U_{j2} h^{-}$. Here $w^{\pm}$ and $h^{\pm}$
are the two-component spinor fields of the wino and the charged
higgsinos, respectively. Furthermore, $U_{ij}$ and $V_{ij}$ are the
elements of the unitary $2\times 2$ matrices, which diagonalize the
chargino mass matrix:
\begin{equation}
M_C=\left(\begin{array}{cc}
M_2 & m_W\sqrt{2}\sin\beta\\
m_W\sqrt{2}\cos\beta & \mu
\end{array}\right)
\hspace*{.1cm}, \label{1N}
\end{equation}
where  $m_W$ denotes the mass of the $W^{\pm}$ bosons.

\subsection{Spin formalism}
We study the production process including the subsequent leptonic
and hadronic decays. { Since in our scenario charginos are very narrow
($\Gamma_{\tilde{\chi}^\pm_1}=2.3$ keV), the narrow-width approximation
is appropriate and
contributions from off-shell channels are negligible;
this approximation can be tested with several
Monte Carlo event generators that include off-shell effects
as well as spin correlations~\cite{whizard}.
The process can therefore be split into the chargino production and
the decay processes.
}
In order to exploit
spin-dependent observables, e.g.\ forward--backward asymmetries
of the final leptons and quarks, { however, the
full spin information of the decaying charginos has to be taken into account.}
In the following we
briefly summarize the required spin formalism. More details as well as
the explicit analytic expressions for the chargino production spin
density matrix and the decay processes can be found
in~\cite{Moortgat-Pick:1998sk,Moortgat-Pick:1997ny}.

The amplitude for the whole process is
\begin{equation}
T=\Delta(\tilde{\chi}^{+}_i) \Delta(\tilde{\chi}^{-}_j) \sum_{\lambda_i,
  \lambda_j}
T_P^{\lambda_i \lambda_j}T_{D, \lambda_i} T_{D, \lambda_j},
\end{equation}
 with the helicity amplitude for the production process
$T_P^{\lambda_i \lambda_j}$ and those for the decay processes
 $T_{D, \lambda_i}$, $T_{D, \lambda_j}$,
 and the propagators $\Delta(\tilde{\chi}^{\pm}_i)=1/[s_i-m_i^2+i
m_i\Gamma_i]$. Here $\lambda_{i}, s_{i}$, $m_{i}$, $\Gamma_{i}$
denote the helicity, four-momentum squared, mass and width of
$\tilde{\chi}^{\pm}_{i}$. The amplitude squared
\begin{equation}
|T|^2 = |\Delta(\tilde{\chi}^{+}_i)|^2 |\Delta(\tilde{\chi}^{-}_j)|^2
\rho^{P,\lambda_i\lambda_j \lambda_i'\lambda_j'}
\rho^{D}_{\lambda_i'\lambda_i}\rho^{D}_{\lambda_j'\lambda_j}
\quad\mbox{(sum convention used)}\label{N}
\end{equation}
 is thus composed of the
(unnormalized) spin density production matrix
\begin{equation}
 \rho^{P,\lambda_i\lambda_j \lambda_i'\lambda_j'}=
T_P^{\lambda_i \lambda_j} T_P^{\lambda_i'\lambda_j' *}
\end{equation}
of $\tilde{\chi}^{\pm}_{i,j}$, and
 the decay matrices
\begin{equation}
\rho^{D}_{\lambda_i'\lambda_i}=
T_{D,\lambda_i} T_{D,\lambda_i'}^{*} \quad\mbox{ and}\quad
\rho^{D}_{\lambda_j'\lambda_j}=
T_{D,\lambda_j} T_{D,\lambda_j'}^{*}.
\end{equation}
 Interference terms between
various helicity amplitudes preclude factorization in a
production factor
$\sum_{\lambda_i \lambda_j} |T_P^{\lambda_i\lambda_j}|^2$
times a decay factor
$\overline{\sum}_{\lambda_i \lambda_j}
|T_{D,\lambda_i\lambda_j}|^2$.

The spin density production matrix
 $\rho^{P, \lambda_i \lambda_j \lambda'_i \lambda'_j}$
can be decomposed into four parts (for details
 see~\cite{Moortgat-Pick:1998sk}):
\begin{equation}
\rho^{P,\lambda_i \lambda_j \lambda'_i \lambda'_j}=
P^{\lambda_i \lambda_j \lambda'_i \lambda'_j}
+\Sigma^{P,\lambda_i \lambda_j \lambda'_i \lambda'_j}_{a}
+\Sigma^{P,\lambda_i \lambda_j \lambda'_i \lambda'_j}_{b}
+\Sigma^{P,\lambda_i \lambda_j \lambda'_i \lambda'_j}
_{ab},\label{24}
\end{equation}
where $P$ denotes a contribution which is
independent of chargino polarizations, $\Sigma_a$ ($\Sigma_b$) depends on the
polarization
of one of the charginos, and  $\Sigma_{ab}$ on both; $a,b=1,2,3$ denote the
components (transverse and longitudinal) of the spin vectors.
Likewise,
the decay matrices
$\rho^D_{\lambda'_i\lambda_i}$
and $\rho^D_{\lambda'_j\lambda_j}$
can each be separated into two parts as
\begin{eqnarray}
\rho^D_{\lambda'_i\lambda_i}&=& D_{\lambda'_i\lambda_i}+
\Sigma^D_{a,\lambda'_i\lambda_i}.\label{24a}
\end{eqnarray}

The amplitude squared $|T|^2$ of the combined processes of production and
decays  can schematically be written as (with helicity indices suppressed):
\begin{eqnarray}
& &|T|^2\sim P D_i D_j +\Sigma^P_a \Sigma^D_a D_j
+\Sigma^P_b \Sigma^D_b D_i +\Sigma^P_{ab} \Sigma_a^D \Sigma_b^D,\label{95}
\end{eqnarray}
The first product in Eq.~(\ref{95}) is independent
of spin correlations between production and decay. The
second and third terms describe the correlations between the
production and the decay process either of $\tilde{\chi}_i^+$ or
$\tilde{\chi}^-_j$ decay and, in the last term correlations between
both decay processes are included.
\begin{itemize}
\item In the first term of Eq.~(\ref{95})
only scalar products appear, which can be expressed
by the Mandelstam variables $s, t, u$
 for the production and decay processes. This is
the only term that survives in the total cross section, i.e.\ when all the
angles are integrated over.
\item In the second (third) term of Eq.~(\ref{95}) the spin vectors relate
  quantities from the
production with those from the decay process. These scalar products cannot be
  expressed
  by Mandelstam
variables. They contain the angle between the incoming electron and
the outgoing lepton or quark in the laboratory system. These terms contribute
  to
spin-dependent observables as, for example, to the forward--backward asymmetry
  of the
  final leptons and quarks.
\item The last term of Eq.~(\ref{95})  involves the spin vectors of both
  charginos and
leads to spin correlations between the two decay chains of $\tilde{\chi}^+_i$
  and   $\tilde{\chi}^-_j$.
\end{itemize}
If the decay of only one chargino, e.g.\ $\tilde{\chi}^+_i$, is
considered, one has to sum over the spin of $\tilde{\chi}^-_j$ so
that in Eq.~(\ref{95}) $D_j=1$ and $\Sigma^D_b=0$.

\subsection{Strategy for the determination of the SUSY parameters}
Our aim is to
demonstrate the power of forward--backward asymmetries in determining
the masses of kinematically inaccessible heavy sleptons. Note, however,
that the leptonic forward--backward asymmetry involves the masses of both
sneutrinos and selectrons. Therefore
the analysis is performed in steps. First we exploit  the chargino and
neutralino masses and the chargino production cross sections times
decay branching fractions, to constrain fundamental parameters of the
chargino and neutralino sectors and the sneutrino mass.
We then show how the obtained limits on the sneutrino mass
and other parameters can be improved by employing the forward--backward
asymmetries, measured
in leptonic chargino decays at the ILC. This, however, requires the
assumption on the $SU(2)$ mass relation for slepton masses. In the last step we
make an attempt at testing the $SU(2)$ symmetry relation for the selectron
and sneutrino, using the available information on the squark masses
from the LHC, and include in addition the forward--backward asymmetry
measured at the ILC in hadronic decay modes.

\section{Case study with heavy sfermions}
\subsection{Parameters of the chosen scenario}

The following mSUGRA parameters, taken at the GUT scale except
for $\tan\beta$, define the MSSM scenario:
\begin{equation}
m_{1/2}=144~\mbox{GeV},\quad m_0=2~\mbox{TeV},\quad
A_0=0~\mbox{GeV}, \quad \tan\beta=20,\quad \mbox{sgn}(\mu)=+1.
\label{sugra-par}
\end{equation}
However, our analysis is performed {\it entirely} within the general
MSSM framework, without any reference to the underlying
SUSY-breaking mechanism. The parameters at the EW scale are obtained
with the help of the SPheno code \cite{Porod:2003um} { for $m_t=178$
GeV}; furthermore it has been checked with the code
micrOMEGA~\cite{Belanger:2004yn} that the lightest neutralino
provides a relic cold-dark-matter density consistent with
cosmological data. The low-scale gaugino/higgsino/gluino masses as
well as the derived masses of SUSY particles are listed in
Tables~\ref{tab_scenario} and \ref{tab_scalar}. The charginos and
neutralinos as well as the gluino are rather light, whereas the
scalar SUSY particles have masses about 2~TeV.

\begin{table}[h]
\begin{center}
\renewcommand{\arraystretch}{1.3}
\begin{tabular}{|c|c|c|c|c||c|c||c|c|c|c||c|}
\hline
$M_1$ & $M_2$ & $M_3$ & $\mu$ & $\tan\beta$& $m_{\tilde{\chi}^{\pm}_1}$
& $m_{\tilde{\chi}^{\pm}_2}$ & $m_{\tilde{\chi}^{0}_1}$  &
$m_{\tilde{\chi}^{0}_2}$ & $m_{\tilde{\chi}^{0}_3}$ & $m_{\tilde{\chi}^{0}_4}$
&
$m_{\tilde{g}}$
\\ \hline
60 & 121 & 322 & 540 &
 20 & 117 & 552 & 59 & 117 & 545 & 550 & 416 \\ \hline
\end{tabular}
\caption{\it Low-scale gaugino/higgsino/$\tan\beta$
MSSM parameters, and the
resulting chargino, neutralino and gluino masses (all masses are given in
GeV).\label{tab_scenario}}
\end{center}
\end{table}

\begin{table}[h]
\begin{center}
\renewcommand{\arraystretch}{1.3}
\begin{tabular}{|c|c|c||c|c|c|c|c||c|c|c|c|}
\hline $m_{h}$ & $m_{H,A}$ & $m_{H^{\pm}}$ & $m_{\tilde{\nu}_e}$ &
$m_{\tilde{e}_{\rm R}}$ & $m_{\tilde{e}_{\rm L}}$ & $m_{\tilde{\tau}_1}$ &
$m_{\tilde{\tau}_2}$ & $m_{\tilde{q}_{\rm R}}$ & $m_{\tilde{q}_{\rm L}}$ &
$m_{\tilde{t}_1}$ & $m_{\tilde{t}_2}$ \\ \hline 119 & 1934 & 1935 &
1994 & 1996 & 1998 & 1930 & 1963 & 2002 & 2008 & 1093 & 1584 \\ \hline
 \end{tabular}
\caption{\it Masses of the SUSY Higgs particles and
scalar SUSY particles (all masses are given in GeV).
\label{tab_scalar}}
\end{center}
\end{table}

\subsection{Expectations at the LHC}
As can be seen from Tables~\ref{tab_scenario} and \ref{tab_scalar}, all
squarks are kinematically accessible at the LHC. The largest squark
production cross  section is for $\tilde{t}_{1,2}$. However, with
stops decaying mainly to $\tilde{g}t$ (with $BR(\tilde{t}_{1,2}\to
\tilde{g} t)\sim 66\%$), where background from top production will
be large, no new interesting channels are open in their decays. The
other squarks decay mainly via $\tilde{g} q$, but since the squarks
are very heavy, $m_{\tilde{q}_{\rm L,R}}\sim 2$~TeV, precise mass
reconstruction will be difficult. Therefore we conservatively
assume that the squark masses can be measured with an error of 50 GeV.
Our results do not depend sensitively
on this assumption since the mere indication that
the scalar quarks are very heavy will be sufficient for narrowing
the experimental uncertainty on the slepton sector from the ILC
measurements.

In this scenario, the inclusive discovery of SUSY at the LHC is
possible mainly because of the large gluino production cross section.
Therefore several gluino decay channels can be exploited. The
largest branching ratio for the gluino decay in our scenario is
a three-body  decay  into
neutralinos, $BR(\tilde{g}\to \tilde{\chi}^0_2 b
\bar{b})\sim 14\%$,
with a subsequent three-body leptonic neutralino decay
$BR(\tilde{\chi}^0_2\to
\tilde{\chi}^0_1 \ell^+ \ell^-)$, $\ell=e,\mu$ of about 6\%, see
Table~\ref{tab:BR}. In this channel the dilepton edge will be clearly
visible, since this process has low
backgrounds~\cite{Weiglein:2004hn}. The mass difference between the
two light neutralinos can be measured from the dilepton edge
with an uncertainty of about \cite{Kawagoe:2004rz}:
\begin{equation}
\delta(m_{\tilde{\chi}^0_2}-m_{\tilde{\chi}^0_1})\sim 0.5~\mathrm{ GeV}.
\label{eq-massdiff}
\end{equation}

\begin{table}[th]
\begin{center}
\renewcommand{\arraystretch}{1.3}
\begin{tabular}{|c||c|c|c|c|c|c|}
\hline Mode &
$\tilde{g}\to \tilde{\chi}^0_2 b \bar{b}$ &
$\tilde{g}\to \tilde{\chi}^{-}_1 q_u \bar{q}_d$ &
$\tilde{\chi}^+_1 \to \tilde{\chi}^0_1 \bar{q}_d q_u$ &
$\tilde{\chi}^0_2\to \tilde{\chi}^0_1 \ell^+\ell^-$ &
$\tilde{t}_{1,2}\to \tilde{g} t$ &
$\tilde{\chi}^-_1 \to \tilde{\chi}^0_1 \ell^- \bar{\nu}_{\ell}$ \\
\hline BR &
$14.4\%$ &
$10.8\%$ &
$33.5\%$ &
$3.0\%$ &
 $66\%$ &
$11.0\%$\\ \hline
\end{tabular}
\caption{\it Branching ratios for some important decay modes in the studied
MSSM scenario, $\ell=e,\mu,\tau$, $q_u=u,c$, $q_d=d,s$. Numbers are given
for each family separately.\label{tab:BR}}
\end{center}
\end{table}

\noindent Other frequent gluino decays  are into the light chargino
and jets, with $BR(\tilde{g}\to \tilde{\chi}^{\pm}_1 q q')\sim 20\%$
for $qq'$ in the first two families, and about $3\%$ in the third,
{ with a subsequent leptonic chargino decay $BR(\tilde{\chi}^\pm_1\to
  \tilde{\chi}^0_1\ell^\pm\nu_\ell), \ell=e,\mu$ of about 11\%.
However,
  exploiting
  this channel for the chargino-neutralino mass difference measurement is
  very
  difficult. First, because of the escaping neutrino,
  and second, because of a genuine 3-body chargino decay in our scenario. To
  our
  knowledge, the only attempt to determine chargino mass at the LHC required
  $\tilde{\chi}^\pm\to \tilde{\chi}^0_1 W^\pm\to
  \tilde{\chi}^0_1\ell^\pm\nu_\ell$ {\it with $W^\pm$ on-shell}  arriving at a
  statistical accuracy of $\sim$25 GeV~\cite{Nojiri:2003tv}.}

{ In both gluino decay channels the spin measurements via angular
  correlations\cite{Barr:2004ze,Wang:2006hk} of decay products should provide
  evidence for the
  spin 1/2   character of the intermediate particles assuring us that the
  underlying SUSY scenario is realized.}

{ Finally, the gluino mass can be reconstructed in a manner similar to
 the one proposed in \cite{Gjelsten:2005aw}, where the SPS1a scenario is
 analyzed.
 Although our scenario is different, the precison in both is limited by
 systematic  uncertainties due to
 hadronic energy scale and a similar relative uncertainty of $\sim$2 \% can be
 expected. }

\subsection{Expectations at the ILC}
At the ILC with $\sqrt{s}\le 500$~GeV, only light charginos and
neutralinos are kinematically accessible. However, in this scenario
the
{ light neutralino sector is characterized by
very low production cross sections. For example, at 500 GeV and
$(P_{e^-},P_{e^+})=(-90\%,+60\%)$ beam polarization we obtain
$\sigma(\tilde{\chi}^0_1\tilde{\chi}^0_2)=0.93$ fb,
$\sigma(\tilde{\chi}^0_2\tilde{\chi}^0_2)=0.49$ fb; for other beam
polarization and/or lower collider energy the cross sections are even
lower. This is due to the almost pure gaugino nature of light neutralinos
($\tilde{\chi}^0_1\sim 99\% \tilde{B}^0,\,
  \tilde{\chi}^0_2\sim 97\% \tilde{W}^0$)
with suppressed couplings to the $s$-channel
$Z$-boson, while the $t$- and $u$-channel selectron exchange is small because
of the heavy selectron mass. Only the $\tilde{\chi}^0_3\tilde{\chi}^0_4$
channel (because of opposite CP
parities of $\tilde{\chi}^0_3$ and $\tilde{\chi}^0_4$)
could have an appreciable
cross section above its threshold $\sqrt{s}\sim 1100$ GeV.}

Only the  chargino pair production process has high rates at the ILC
and all information obtainable from this sector has to be used.
We constrain our analysis to the first stage of the ILC with
$\sqrt{s} \le 500$~GeV and study only the
$\tilde{\chi}^+_1\tilde{\chi}^-_1$ production
\begin{equation}
e^+ e^- \to \tilde{\chi}^+_1 \tilde{\chi}^-_1, \label{eq:prod}
\end{equation}
with subsequent chargino decays
\begin{equation}
\tilde{\chi}^-_1\to \tilde{\chi}^0_1 e^- \bar{\nu}_e, \;
\tilde{\chi}^0_1 \mu^- \bar{\nu}_\mu,\;
\tilde{\chi}^0_1 d \bar{u} \quad{\rm and}\quad
\tilde{\chi}^0_1 s \bar{c} \label{eq:dec}
\end{equation}
and the corresponding charge conjugate $\tilde{\chi}^+_1$ decays,
for which the analytical formulae, including the complete spin
correlations, are given in a compact form, e.g.\
in~\cite{Moortgat-Pick:1998sk}.

\subsubsection{3.3.1 Mass measurements}
The chargino mass can be measured at $\sqrt{s}=350$ and $500$~GeV in
the continuum, with an error of about $ 0.5$
GeV~\cite{Martyn:1999tc,Accomando:1997wt}. This can serve to
optimize the ILC scan at the threshold~\cite{Stirling:1995xp} which,
because of the steepness of the $s$-wave excitation curve in
$\tilde{\chi}^+_1\tilde{\chi}^-_1$ production, can be used to
determine the light chargino mass very precisely, to
about~\cite{Accomando:1997wt}:
\begin{equation}
m_{\tilde{\chi}^{\pm}_1}=117.1 \pm 0.1 ~\mathrm{ GeV}.
\label{eq_massthres}
\end{equation}

The mass of the lightest neutralino
$m_{\tilde{\chi}^0_1}$ can be derived via the decays of the light chargino,
either from the energy
distribution of the lepton $\ell^-$ ($BR(\tilde{\chi}^{-}_1\to
\tilde{\chi}^0_1 \ell^-
\bar{\nu}_\ell)\sim 11\%$, see Table~\ref{tab:BR})
or from the invariant mass distribution of the two jets
in  hadronic decays ($BR(\tilde{\chi}^-_1\to \tilde{\chi}^0_1 q_d
\bar{q}_u)\sim 33\%$, see Table~\ref{tab:BR}). We
  take~\cite{Accomando:1997wt}
\begin{equation}
m_{\tilde{\chi}^{0}_1}=59.2 \pm 0.2~\mathrm{ GeV}.
\label{eq_masslsp}
\end{equation}
Together with the information from the LHC, Eq.~(\ref{eq-massdiff}),
a mass uncertainty for the second lightest neutralino of about
\begin{equation}
m_{\tilde{\chi}^{0}_2}=117.1 \pm 0.5 ~\mathrm{ GeV}
\label{eq_masschi02}
\end{equation}
can be assumed.

\subsubsection{3.3.2 Experimental uncertainties for the cross sections}
Table~\ref{tab_cross} lists the expected chargino production cross
sections and forward--backward asymmetries for different beam
energies and polarization configurations, derived with the nominal
values of parameters. Experimentally we identify the chargino pair
production process, $e^+e^- \to \tilde{\chi}_1^+ \tilde{\chi}_1^-$,
in the fully leptonic ($\ell^+\nu\tilde{\chi}^0_1\ell^-
\bar{\nu}\tilde{\chi}^0_1$) and semileptonic
($\ell\nu\tilde{\chi}^0_1q\bar{q}'\tilde{\chi}^0_1$) final states
(where $\ell=e,\mu$). We estimate an overall selection efficiency of
50\%. For both final states, $W^+W^-$ production is expected to be
the dominant SM background. The fully leptonic channel is more
challenging, owing to the absence of mass constraints. Its efficient
selection needs further experimental study. However, the fully
leptonic channel is not essential for this analysis, as its relative
contribution to the overall rate is only $\sim 14$\%. For the
semileptonic (slc) final state, this background can be efficiently
reduced from the reconstruction of the hadronic invariant mass.  In
Table~\ref{tab_cross}, we list cross sections multiplied by the
branching fraction $B_{slc}=2\times BR(\tilde{\chi}^+_1 \to
\tilde{\chi}^0_1 \bar{q}_d q_u)\times BR(\tilde{\chi}^-_1 \to
\tilde{\chi}^0_1 \ell^- \bar{\nu})+ [BR(\tilde{\chi}^-_1 \to
\tilde{\chi}^0_1 \ell^- \bar{\nu})]^2\sim 0.34$ ($\ell,\, q_d,\,
q_u$ include the first two families of leptons and quarks) including an
$e_{slc}=50\%$ selection efficiency. The error includes, added in
quadrature, the statistical uncertainty based on numbers of identified
events for ${\cal L}=200$~fb$^{-1}$ in each polarization
configuration, $(P_{e^-},P_{e^+})=(-90\%,+60\%)$ and
$(+90\%,-60\%)$, and a relative uncertainty in the polarization of
$\Delta P_{e^\pm}/P_{e^\pm}=0.5\%$~\cite{Moortgat-Pick:2005cw}.  Since
the production rates are high, the total uncertainties are rather
small; see Table~\ref{tab_cross}.

\begin{table}[h]
\begin{center}
\renewcommand{\arraystretch}{1.3}
\begin{tabular}{|l c||r|r||r|r|} \hline $\sqrt{s}$/GeV &
$(P_{e^-},P_{e^+})$ & $\sigma(\tilde{\chi}^+_1\tilde{\chi}^-_1)$/fb
& $\sigma(\tilde{\chi}^+_1\tilde{\chi}^-_1)\, B_{slc}\, e_{slc} $/fb
& $A_{\rm FB}(\ell^-)$/\% & $A_{\rm FB}(\bar{c})$/\% \\
\hline\hline
350 & $(-90\%,+60\%)$ & 6195.5
    & 1062.5$\pm$4.0 & {4.42$\pm$0.29} & 4.18$\pm$0.74 \\
\hline & $(+90\%,-60\%)$ & 85.0
     & 14.6$\pm 0.7$ & -- & --\\
\hline\hline 500 & $(-90\%,+60\%)$ & 3041.5
     & 521.6$\pm 2.3$ & {4.62$\pm$0.41} & 4.48$\pm1.05$\\
\hline & $(+90\%,-60\%)$ & 40.3
     & 6.9$\pm 0.4$ & -- & --\\
\hline
\end{tabular}
\caption{\it Cross sections for the process $e^+e^-\to
\tilde{\chi}^+_1\tilde{\chi}^-_1$ and forward--backward asymmetries ($A_{\rm
  FB}$)
in the leptonic $\tilde{\chi}^-_1 \to \tilde{\chi}^0_1 \ell^-
\bar{\nu}$ and hadronic $\tilde{\chi}^-_1 \to \tilde{\chi}^0_1 s
\bar{c}$ decay modes, for different beam polarization $P_{e^-}$,
$P_{e^+}$ configurations at centre-of-mass energies $\sqrt{s}=350$~GeV and
$500$~GeV at the ILC. Concerning the errors, see text.
\label{tab_cross}}
\end{center}
\end{table}

{\subsubsection{3.3.3 Experimental uncertainties for the forward--backward
    asymmetries}

Figure~\ref{fig-spinc} shows the expected forward--backward asymmetries
measured in the leptonic and hadronic decay channels as functions of
the sneutrino mass; { for illustration, the dashed line in the left panel
  demonstrates  that spin correlations between production and decay must be
  taken into account
  for a proper interpretation of the experimental data}. In the case of
leptonic decay the $SU(2)$
relation on slepton masses
\begin{equation}
m^2_{\tilde{e}_{\rm L}}= m_{\tilde{\nu}_e}^2 +
m_Z^2 \cos(2 \beta) (-1+\sin^2\theta_W)
\label{eq:su2}
\end{equation}
has been assumed, while for the hadronic decay
the squark masses are taken to be fixed by the LHC
measurement.

The forward--backward asymmetry is defined as:
\begin{eqnarray}
A_{\rm FB}
= \frac{\sigma_{\rm F} - \sigma_{\rm B}}{\sigma_{\rm F} + \sigma_{\rm B}},
\end{eqnarray}
where $\sigma_{\rm F}$
and $\sigma_{\rm B}$ are the acceptance-corrected cross sections.
The statistical error on $A_{\rm FB}$, { assuming a binomial distribution,}
is given by
\begin{equation}
\Delta(A_{\rm FB})^{\mathrm{stat}}=2 \sqrt{\epsilon (1-\epsilon)/N},
\label{error-afb}
\end{equation}
where $\epsilon=\sigma_{\rm F}/(\sigma_{\rm F}+\sigma_{\rm B})$ and $N$
denotes the
number of selected events. { Errors due to uncertainties of beam
  polarization and squark masses are negligible.}

For the {\it leptonic} forward--backward asymmetry, $\sigma_{\rm F}$
and $\sigma_{\rm B}$ are the acceptance-corrected cross sections for the
$e^-/\mu^-$ from $\tilde{\chi}^-_1$ being observed in the forward (F) and
backward (B)
hemisphere, respectively, for $\ell^- \bar{\nu} \tilde{\chi}^0_1 q
\bar{q}' \tilde{\chi}^0_1$ and $\ell^- \bar{\nu} \tilde{\chi}^0_1
\ell^+ \nu \tilde{\chi}^0_1$ final states. For the $\ell^+ \nu
\tilde{\chi}^0_1 q \bar{q}' \tilde{\chi}^0_1$ final state, in which $\ell^+$
comes from $\tilde{\chi}^+_1$ decay, the
condition is that the positive lepton is observed in the backward
(F) or forward (B) direction.

The {\it hadronic} forward--backward asymmetry has been analyzed
if one flavour of the hadronic 2-jet system can be identified. Since
the expected vertex-detector performance is excellent at the ILC,
charm-tagging from secondary vertices and displaced tracks can be
envisaged. This allows us to measure the forward--backward asymmetry of the
$c/{\bar{c}}$-jet in each of the different semileptonic decays $\ell^+ \nu
\tilde{\chi}^0_1 \bar{c} s
\tilde{\chi}^0_1$ and $\ell^- \bar{\nu} \tilde{\chi}^0_1 c \bar{s}
\tilde{\chi}^0_1$. Note that the charge sign of the $c$-jet
can be inferred from the lepton charge of the other chargino decay and does
not have to be
measured through jet- or vertex-charge techniques.  Taking the
results from \cite{Hillert:2005ss}, $c$-jets can be identified with
an efficiency of 50\% at a purity of 80\% in $Z$-decays at rest. The
purity from chargino and $W$ decays is larger, since the ratio of
true charm/non-charm jets is $\sim 1/3$ (compared to $\sim 1/5$ for
$Z$ decays). In this analysis, we assume an overall selection
efficiency for $\ell \nu \tilde{\chi}^0_1 c s \tilde{\chi}^0_1$ of
20\%, corresponding to a $c$-tag efficiency of 40\% for a selection
efficiency of 50\%. In Table~\ref{tab_cross} the asymmetries are
listed only for the $(P_{e^-},P_{e^+})=(-90\%,+60\%)$ case, since the cross
sections for
the opposite polarization are very small and the
statistical errors become
very large. Consequently we do not include them in the following analysis.

\begin{figure}
\setlength{\unitlength}{1cm}
\begin{center}
\begin{picture}(15,6.5)
\put(1.7,6.2){\small $e^+e^-\to \tilde{\chi}^+_1 \tilde{\chi}^-_1$,
$\tilde{\chi}^-_1 \to \tilde{\chi}^0_1 \ell^- \bar{\nu}$}
\put(0.1,.2){\mbox{\epsfig{figure=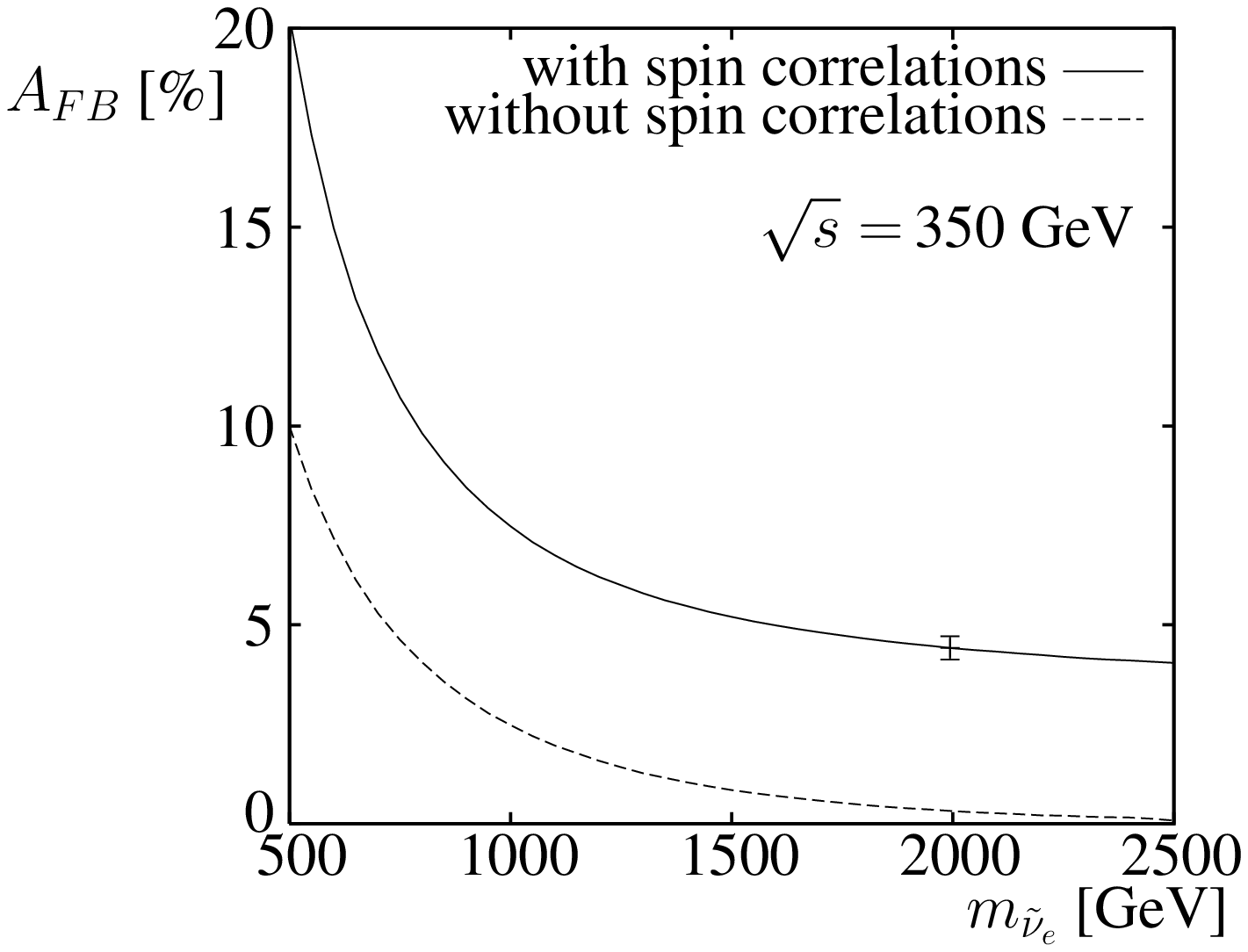,height=5.8cm,width=7cm}}}
\put(8.2,.2){\mbox{\epsfig{figure=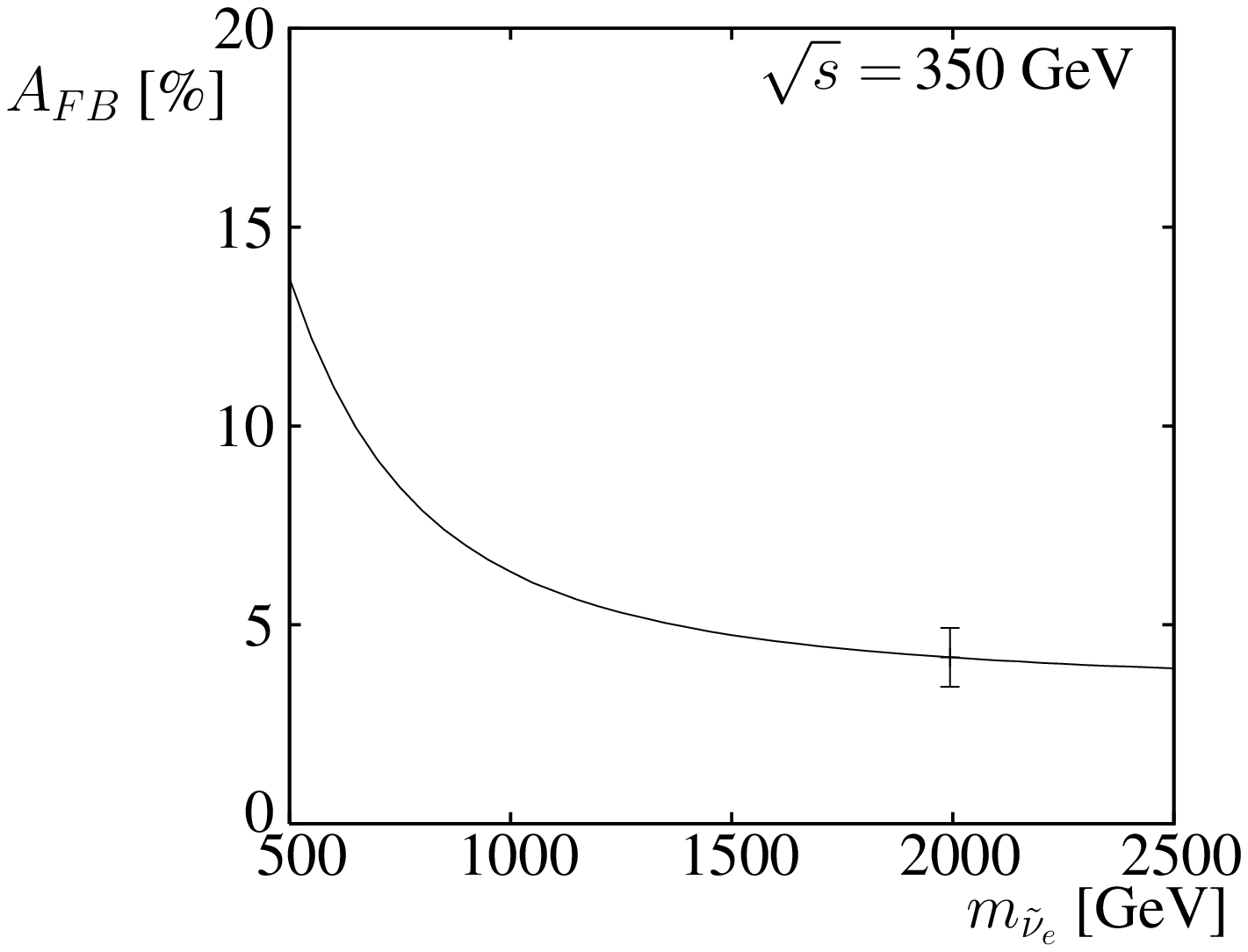,
height=5.8cm,width=7cm}}}
\put(9.8,6.2){\small $e^+e^-\to\tilde{\chi}^+_1 \tilde{\chi}^-_1$,
  $\tilde{\chi}^-_1 \to
\tilde{\chi}^0_1 s \bar{c}$}
\end{picture}
\end{center}
\vspace{-.8cm} \caption{\it Forward--backward asymmetry of $e^-$ in
the process $e^+e^- \to \tilde{\chi}^+_1 \tilde{\chi}^-_1$,
$\tilde{\chi}^-_1 \to \tilde{\chi}^0_1 \ell^- \bar{\nu}$ (left
panel) and of $\bar{c}$-jet in the  process $e^+e^- \to
\tilde{\chi}^+_1 \tilde{\chi}^-_1$, $\tilde{\chi}^-_1 \to
\tilde{\chi}^0_1 s \bar{c}$ (right panel), shown as a function of
$m_{\tilde{\nu}_e}$ at $\sqrt{s}=350$~GeV and polarized beams
$(P_{e^-},P_{e^+})=(-90\%,+60\%)$. In the left panel the mass of the
other scalar virtual particle, $m_{\tilde{e}_{\rm L}}$, which contributes
to the decay process, has been assumed to fulfill the $SU(2)$ mass
relation, Eq.~(\ref{eq:su2}), while in the right panel the mass of
the squark is kept fixed as measured at the LHC.
For nominal value of $m_{\tilde{\nu}_e}=1994$ GeV the
expected experimental errors are shown, { see Eq.~\ref{error-afb}.  For
  illustration only, the dashed
line in the left panel shows that neglecting spin
correlations would lead to a completely wrong interpretation of the
experimental data.}
\label{fig-spinc}}
\end{figure}

\section{Parameter determination}

In the following we apply multiparameter fits to determine the underlying
SUSY parameters.
\begin{itemize}
\item In the first step we use only
the masses $\tilde{\chi}^{\pm}_1$, $\tilde{\chi}^0_1$,
$\tilde{\chi}^0_2$ and the chargino pair production cross section,
including the full leptonic and the semileptonic decays as
observables. A four-parameter fit for the parameters $M_1$, $M_2$,
$\mu$ and $m_{\tilde{\nu}}$ has been applied.
\item In the second step we include as an additional observable the leptonic
  forward--backward
asymmetry. Only the semileptonic and purely leptonic decays were
used. The $SU(2)$ relation between the two virtual masses
$m_{\tilde{\nu}}$ and
  $m_{\tilde{e}_{\rm L}}$ has
been applied as an external constraint.
\item As an attempt to test the $SU(2)$ mass relation for the slepton and
sneutrino
  masses, in the last step both the leptonic and hadronic forward--backward
  asymmetries have been used.
A six-parameter fit for the parameters $M_1$, $M_2$, $\mu$, $m_{\tilde{\nu}}$,
$m_{\tilde{e}_{\rm L}}$ and $\tan\beta$
has been  applied.
\end{itemize}

\subsection{Parameter fit without using the forward--backward asymmetry}
\label{sec:woafb}
We use as observables the masses $m_{\tilde{\chi}^{\pm}_1}$,
$m_{\tilde{\chi}^0_1}$,
$m_{\tilde{\chi}^0_2}$ and the polarized chargino cross section multiplied by
the
branching ratios of semileptonic chargino decays:
$\sigma(e^+e^-\to\tilde{\chi}^+_1\tilde{\chi}^-_1)\times B_{slc}$, with
$B_{slc}=2\times BR(\tilde{\chi}^+_1 \to \tilde{\chi}^0_1 \bar{q}_d
q_u)\times BR(\tilde{\chi}^-_1 \to \tilde{\chi}^0_1 \ell^-
\bar{\nu})+ [BR(\tilde{\chi}^-_1 \to \tilde{\chi}^0_1 \ell^-
\bar{\nu})]^2\sim 0.34$, $\ell=e,\mu $, $q_u=u,c$, $q_d=d,s$, with
selection efficiency 50\%, as given in Table~\ref{tab_cross}. Note
that the chargino branching ratios are not sensitive functions of
sfermion masses mediating their decays, since we know from the LHC
that sfermions are very heavy. We take into account the $1\sigma$
statistical error based on ${\cal L}=200$~fb$^{-1}$ for each
polarization configuration, a relative uncertainty in polarization
of $\Delta
P_{e^{\pm}}/P_{e^{\pm}}=0.5\%$~\cite{Moortgat-Pick:2005cw} and an
experimental efficiency of 50\%; see Table~\ref{tab_cross}.

We apply a four-parameter fit for the parameters $M_1$, $M_2$,
$\mu$ and $m_{\tilde{\nu}_e}$ for fixed values of $\tan\beta=5$, 10, 15, 20,
25, 30, 50 and 100. Fixing $\tan\beta$ is necessary for a
proper convergence of the fitting
{ procedure
  \cite{James:1975dr}
  because of strong correlations among parameters}. We perform a
$\chi^2$ test
\begin{equation} \label{eq:chi2}
    \chi^2=\sum_{\substack{i={\rm LR}, {\rm RL}\\j=350,500}}
    \Big(\frac{\sigma(ij)-\sigma(ij)^{\mathrm{th}}}
{\Delta\sigma(ij)}\Big)^2+
    \sum_{i=\tilde{\chi}^\pm_1,\tilde{\chi}^0_1,\tilde{\chi}^0_2}
    \Big(\frac{m_i-m_i^{\mathrm{th}}} {\Delta
    m_i}\Big)^2,
\end{equation}
where in the first term $i={\rm LR}, {\rm RL}$ denote $(e^-,e^+)$
polarization configurations $(-90\%,+60\%)$ and $(+90\%,-60\%)$
respectively, and $j=350, 500$ denote the c.m.\ energy in GeV;
$\sigma(ij),\, m_i$ and $\Delta\sigma(ij),\,\Delta m_i$ are the
corresponding cross sections and masses and their uncertainties; see
Table~\ref{tab_cross} and
Eqs.~(\ref{eq_massthres})--(\ref{eq_masschi02}).

\begin{figure}
\setlength{\unitlength}{1cm}
\begin{center}
\begin{picture}(15,5)
\put(0.8,4.7){\small $\tan\beta=$} \put(-1.0,4.7){\small $M_2$}
\put(3.8,-.1){\small $m_{\tilde{\nu}_e}$} \put(2.5,4.18){\small 5}
\put(2.65,2.75){\small 10} \put(2.67,1.95){\small 20}
\put(2.71,1.4){\small 50}
\put(-0.4,.2){\mbox{\epsfig{figure=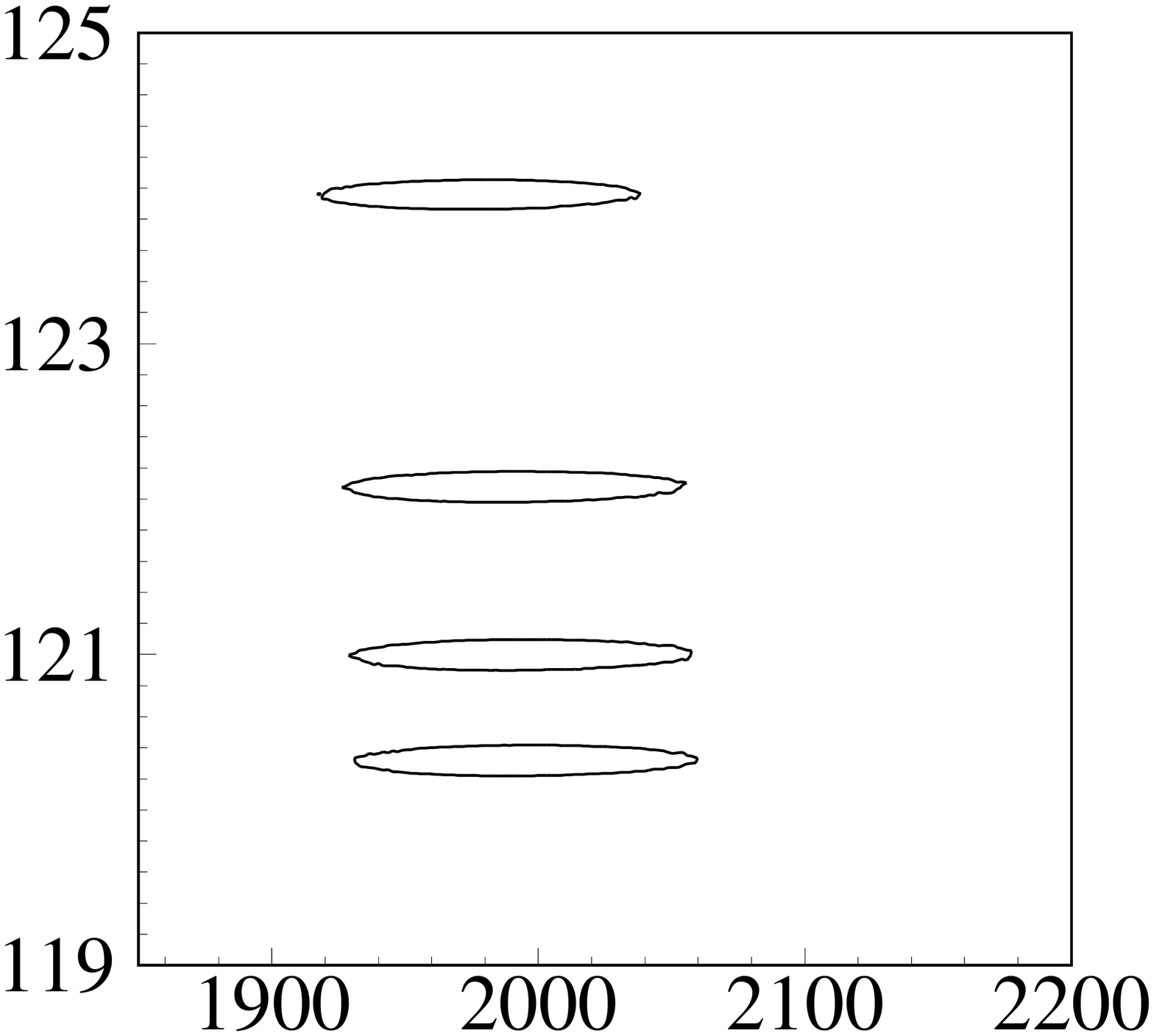,height=5.cm,width=5cm}}}
\put(4.7,4,7){\small $\mu$} \put(9.2,-.2){\small $M_2$}
\put(6.2,4.7){\small $\tan\beta=$} \put(8.60,4.16){\small 5}
\put(7.2,3.7){\small 10} \put(6.5,3.47){\small 20}
\put(5.95,3.35){\small 50}
\put(5.,.2){\mbox{\epsfig{figure=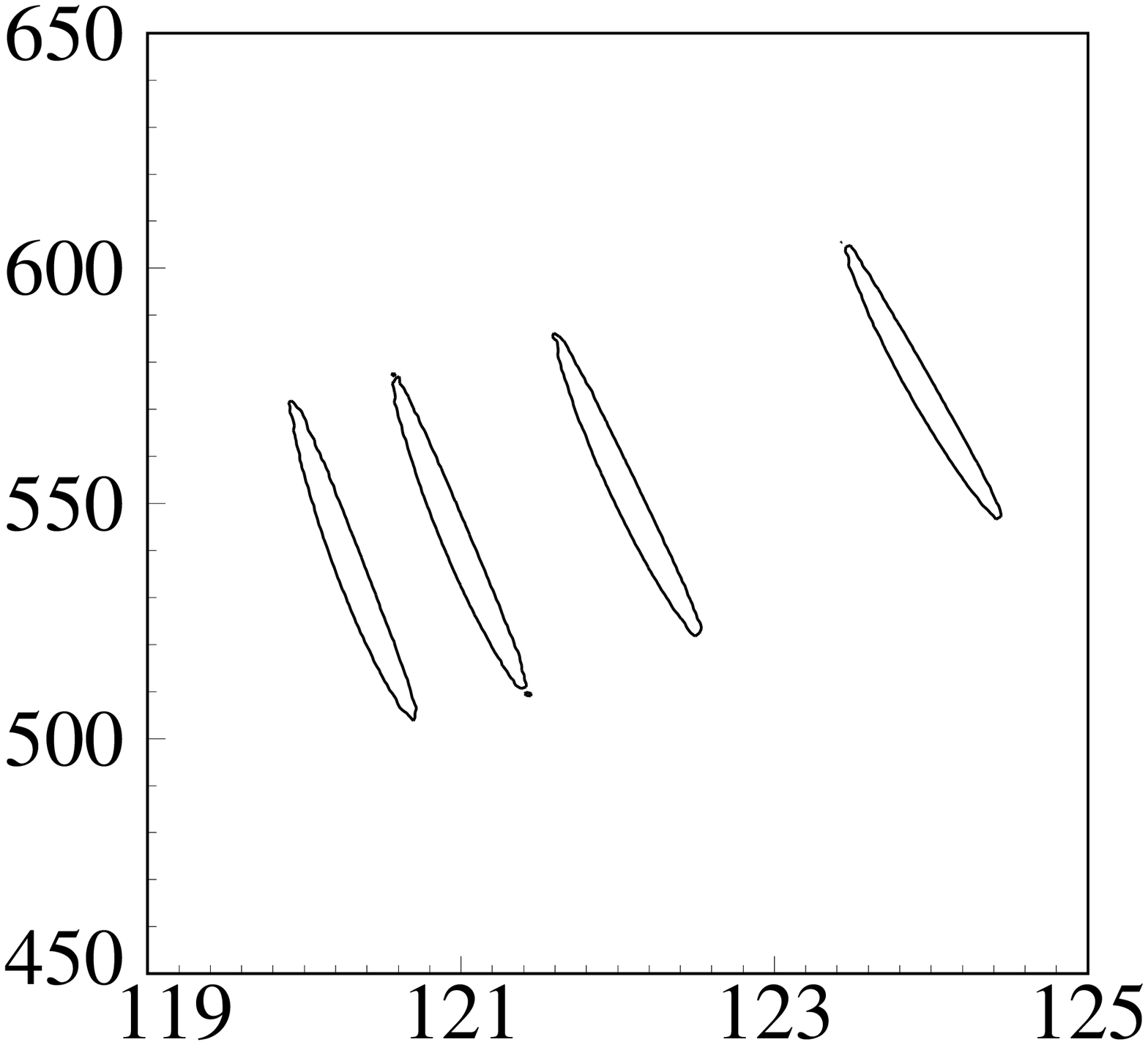,height=5.cm,width=5cm}}}
\put(10.1,4.7){\small $M_2$} \put(14.9,-.2){\small $M_1$}
\put(11.8,4.7){\small $\tan\beta=$} \put(14.78,4.18){\small 5}
\put(13.5,2.75){\small 10} \put(12.8,1.92){\small 20}
\put(12.35,1.4){\small 50}
\put(10.7,.2){\mbox{\epsfig{figure=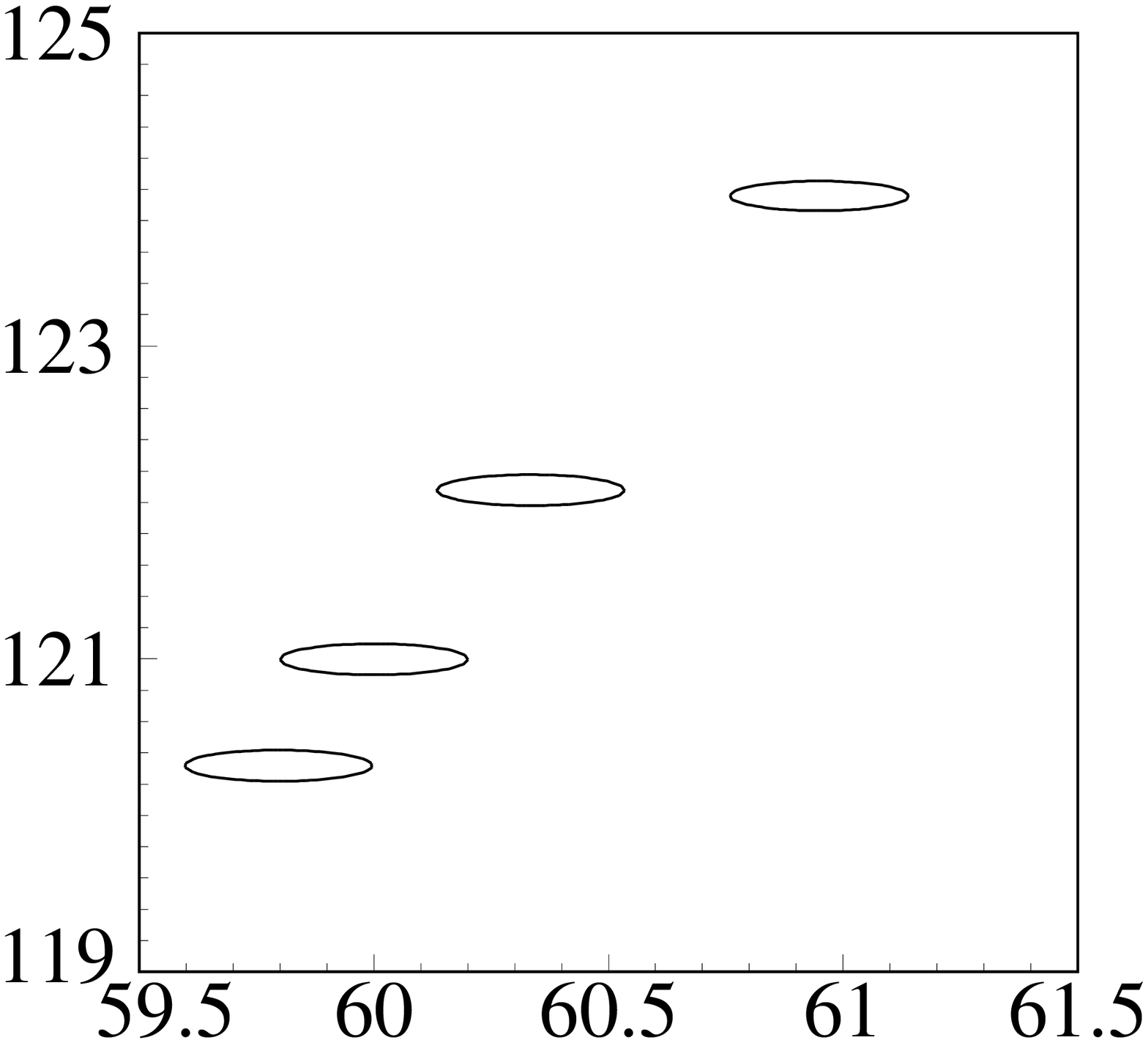,height=5.cm,width=5cm}}}
\end{picture}
\end{center}
\caption{\it Migration of 1$\sigma$ contours for $\tan\beta=5,\,
10,\, 20,\, 50$ with the other two parameters fixed at the values
determined by the minimum of $\chi^2$ for each $\tan\beta$.
\label{migration}}
\end{figure}
It turns out that  for $\tan\beta<1.7$
the measurements are inconsistent with theoretical predictions at
least at the 1$\sigma$ level. The corresponding
1$\sigma$ constraints from cross sections and light neutralino and
chargino mass measurements for the underlying parameters are as
follows
\begin{eqnarray}
&&59.4\le M_1\le 62.2~\mathrm{ GeV},\quad
118.7\le M_2\le 127.5~\mathrm{ GeV}, \nonumber \\
&&450\le \mu\le 750~\mathrm{ GeV},\quad
1800\le m_{\tilde{\nu}_e}\le 2210~\mathrm{ GeV}.
\label{eq:xesct}
\end{eqnarray}
Owing to the strong gaugino component of $\tilde{\chi}^{\pm}_1$ and
$\tilde{\chi}^0_{1,2}$, the parameters $M_1$ and $M_2$ are
determined reasonably well, with a relative uncertainty of $\sim 3\%$ and
$\sim 5\%$. The higgsino parameter $\mu$ as well as
$m_{\tilde{\nu}_e}$ are determined to a lesser degree of precision, with
relative
errors of $\sim 30\%$ and 10\%. Note, however, that large errors are
partly due to migration of the fitted central values of the
parameters  with $\tan\beta$. Figure~\ref{migration} shows the
migration of 1$\sigma$ contours\footnote{Note that
  these plots are 2-dim cuts of a 4-dim hypersurface for each $\tan\beta$ value
  and they may give a false impression that errors are smaller than that in
  Eq.~(\ref{eq:xesct}).} in $m_{\tilde{\nu}_e}$--$M_2$
(left), $M_2$--$\mu$ (middle)  and $M_1$--$M_2$ (right), the
other two parameters being fixed at the values determined by the minimum
of $\chi^2$ for $\tan\beta$ changing from 5 to 10, 20 and 50. Beyond
$\tan\beta=50$, the migration is negligible. Varying $\tan\beta$
between 5 and 50 leads to a shift $\sim 1$~GeV of the fitted central
$M_1$ value and $\sim 3.5$ GeV of $M_2$, effectively increasing
their experimental errors, while the migration effect for $\mu$ and
$m_{\tilde{\nu}_e}$ is much weaker.
%

\subsection{Parameter fit including the leptonic forward--backward\\ asymmetry}
\label{sec:with}
We now extend the fit by using as
additional observable the leptonic forward--backward asymmetry for
polarized beams $(-90\%,+60\%)$. We include final-state electrons
and muons, assuming equal masses of selectrons and smuons, and we
include decays of both charginos. The $SU(2)$ relation
between selectron and sneutrino masses has been assumed, see
Eq.~(\ref{eq:su2}). The parameter ranges found in the previous step
are scanned and accepted if $\chi^2_{A_{\rm FB}}\le 1$ after inclusion
of forward--backward asymmetry according to
\begin{equation}
\chi^2_{A_{\rm FB}}=\chi^2 + \sum_{i}
    \Big(\frac{A_{\rm FB}(i)-A_{\rm FB}(i)^{\mathrm{th}}}{\Delta
    A_{\rm FB}(i)}\Big)^2,
\label{eq:chisq}
\end{equation}
where $\chi^2$ is defined as in Eq.~(\ref{eq:chi2}), and  the sum
over $i$ includes $A_{\rm FB}$ measured for both electrons and muons
at c.m.\ energies of 350 and 500 GeV. The terms $A_{\rm FB}(i)$ and
$\Delta A_{\rm FB}(i)$ are the corresponding experimental
forward--backward asymmetries and their uncertainties; see
Table~\ref{tab_cross}. For the forward--backward asymmetries the
errors due to the uncertainty of beam polarization, although very
small with respect to the statistical one,  are also included in the
$\chi^2$ test.
\begin{figure}
\setlength{\unitlength}{1cm}
\begin{center}
\begin{picture}(15,5)
\put(0.4,.2){\mbox{\epsfig{figure=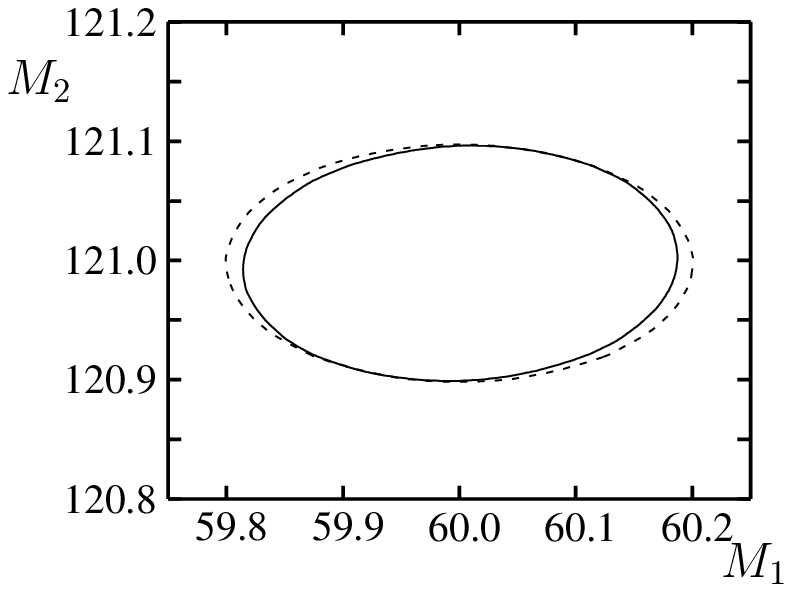,height=5.cm,width=6cm}}}
\put(7.,.2){\mbox{\epsfig{figure=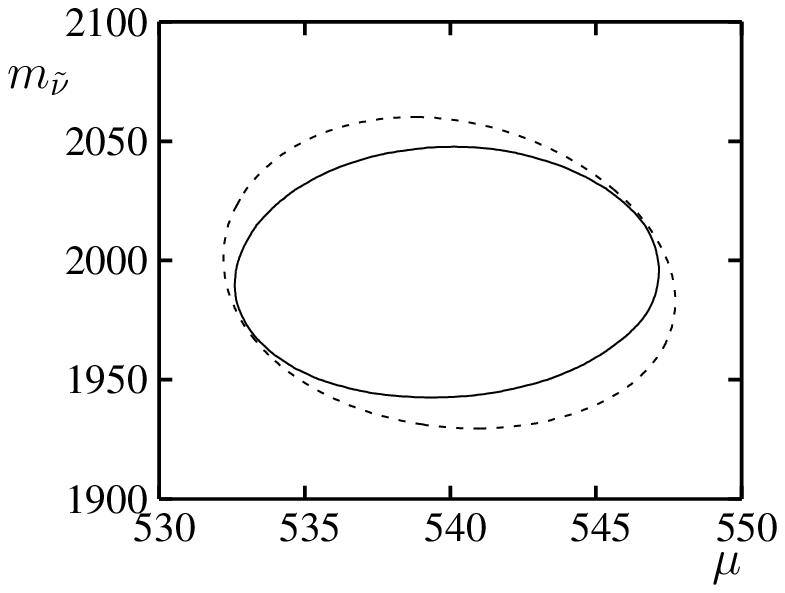,height=5.cm,width=6cm}}}
\end{picture}
\end{center}
\caption{\it  Two 2-dim cuts in \{$M_1$-$M_2$\} and \{$\mu$-$m_{\tilde{\nu}}$\}
  planes of the 1$\sigma$ hypersurface
before (dashes) and after (solid) including the leptonic $A_{\rm FB}$
  in the fit for fixed $\tan\beta=20$ and the other
parameters taken at the values
determined by the minimum of $\chi^2$.
\label{afterAfb}}
\end{figure}
{ The effect of including the leptonic FB asymmetry in the fit is
  illustrated in
Fig.~\ref{afterAfb} for two 2-dim cuts of the 1$\sigma$ hypersurface at {\it
  fixed}
$\tan\beta=20$. As expected, the range of $m_{\tilde{\nu}}$ is most affected,
  although changes of contours are not large.
However, the main virtue of including $A_{\rm FB}$, not visible in this Figure,
is {\it constraining} $\tan\beta$. No assumption on
$\tan\beta$ has to be made in the fit since for too small or too large
a value of $\tan\beta$ the wrong
value of $A_{\rm FB}$ is predicted. As a result,
including the forward--backward  asymmetries in the multiparameter
fit strongly improves the results. We find
\begin{eqnarray}
&&59.7\le M_1\le 60.35~\mathrm{ GeV},\quad 119.9\le M_2 \le
122.0~\mathrm{ GeV},\quad 500\le \mu\le 610~\mathrm{ GeV},\nonumber
\\ && 1900\le m_{\tilde{\nu}_e}\le 2100~\mathrm{ GeV},\quad 14\le
\tan\beta \le 31. \label{eq:leptonic}
\end{eqnarray}
In particular,  $\tan\beta$ is constrained from
below rather well.}
 The constraints for the mass $m_{\tilde{\nu}_e}$
are improved by a factor of about $2$ and for gaugino mass
parameters $M_1$ and $M_2$ by a factor of about $5$, as compared to the
results of Section~\ref{sec:woafb} with unconstrained
$\tan\beta$. The error for the higgsino mass parameter $\mu$ also
decreases significantly.
{ From Eq.~(\ref{eq:leptonic}) we obtain the following predictions for the
  heavy  charginos/neutralinos
\begin{eqnarray}
506 \le m_{\tilde{\chi}^0_3}\le 615 {\rm ~GeV},\quad
512 \le m_{\tilde{\chi}^0_4} \le 619 {\rm ~GeV},\quad
514\le m_{\tilde{\chi}^{\pm}_2}\le 621 {\rm ~GeV}.\label{eq_mchi2_pred}
\end{eqnarray}
These sparticles are only kinematically accessible with non-negligible
cross sections at a phase-2 ILC.}


\subsection{Parameter fit including the hadronic and leptonic\\
forward--backward asymmetries: test of $SU(2)$}

\subsubsection{4.3.1 Parameter fit including the leptonic $A_{\rm FB}$}
In principle the $SU(2)$ relation can be tested by employing the
forward--backward asymmetries measured in hadronic and leptonic
decay modes of produced chargino. With the constraints for the
squark masses from the LHC, the hadronic forward--backward asymmetry
could be used to control the sneutrino mass and the leptonic
forward--backward asymmetry to derive constraints on the selectron
mass. However, with the foreseen experimental accuracies, testing
the $SU(2)$ relation turns out to be very challenging { because our
measurements do not sufficiently constrain all 6 parameters
  simultaneously. Therefore,}
we perform a scan of the parameter space
and calculate $\chi^2_{A_{\rm FB}}$ according to Eq.~(\ref{eq:chisq}), i.e.\
taking light chargino and neutralino masses
(Eqs.~(\ref{eq_massthres}-\ref{eq_masschi02})), and 4 cross section
measurements and 2 leptonic forward--backward asymmetry measurements
(entries in columns 3 and 4 of Table~\ref{tab_cross}). From
$\chi^2_{A_{\rm FB}}=1$ we derive the  following constraints:
\begin{eqnarray}
&&59.30\le M_1\le 60.80~\mathrm{ GeV},\quad 117.8\le M_2\le
124.2~\mathrm{ GeV}, \quad 420\le \mu\le 950~\mathrm{ GeV},
\nonumber
\\ \quad && 1860\le m_{\tilde{\nu}_e}\le 2200~\mathrm{ GeV},
\quad m_{\tilde{e}_{\rm L}}\ge 1400~\mathrm{GeV}, \quad \tan\beta\ge 11.
\label{eq:nosu2}
\end{eqnarray}
{ As we can see, without  the $SU(2)$ relation the upper limits on the
  selectron mass and $\tan\beta$ cannot be established
since a change of these
parameters for high values can be compensated by small changes of
other parameters.}
Limits for the parameter $\mu$ are also very poor.
The parameters $M_1$ and $M_2$ are nevertheless quite well
determined, with an accuracy of the order of a few per cent,
thanks to tight experimental mass constraints on the light chargino and
neutralinos.

\subsubsection{4.3.2 Parameter fit including the leptonic and hadronic
  $A_{\rm FB}$}
Including hadronic forward--backward asymmetry (two entries in the last column
of Table~\ref{tab_cross}) improves
the constraints as follows:
\begin{eqnarray}
&&59.45\le M_1\le 60.80~\mathrm{ GeV},\quad 118.6\le M_2\le
124.2~\mathrm{ GeV}, \quad 420\le \mu\le 770~\mathrm{ GeV},
\nonumber
\\ \quad && 1900\le m_{\tilde{\nu}_e}\le 2120~\mathrm{ GeV},
\quad m_{\tilde{e}_{\rm L}}\ge 1500~\mathrm{GeV}, \quad
11\le\tan\beta\le 60. \label{eq:hadnosu2}
\end{eqnarray}
The most significant change is for the sneutrino mass, for which error bars
become smaller by $\sim 50\%$. Also an upper limit on $\tan\beta$ is found,
which has the effect of improving the upper limit on $\mu$ significantly and
slightly
lowering the limits for $M_1$ and $M_2$.
However we do not get an upper limit for the selectron mass.
Nevertheless, the results  for the
selectron and sneutrino masses are consistent with the $SU(2)$ relation.
The hadronic forward--backward
asymmetry would be much more
useful with more precise measurements, which is very challenging
experimentally.

\section{Conclusions}

We showed a method for determining the MSSM parameters in scenarios
with heavy scalar particles where only a small part of the particle
spectrum is kinematically accessible at the ILC. Such scenarios
appear very challenging, since only a very limited amount of
experimental information is accessible about the SUSY sector. However,
a careful exploitation of data leads to significant constraints for
unknown parameters. A very powerful tool in this kind of analysis
turns out to be the forward--backward asymmetry. The proper treatment
of spin correlations between the production and the decay is
necessary in this context. This asymmetry is strongly dependent on
the mass of the exchanged heavy particle. If the $SU(2)$ constraint
is applied, the slepton masses can be determined to a precision of
about 5\% for masses around 2~TeV at the ILC running at 500~GeV,
i.e.\ one eighth of the energy necessary for slepton pair
production. Also the derived constraints on heavy chargino/neutralinos may
provide the physics argument for a second stage of the ILC.

The $SU(2)$ assumption on the left-selectron and sneutrino masses
could in principle be tested by combining the leptonic forward--backward
asymmetry with the hadronic forward--backward asymmetry
for which the squark masses are measured at the LHC. However,
with current estimates for the efficiency and purity of charm tagging at the
ILC, this test is not very stringent. With significantly better charm-tagging
performance more sensitive tests could be performed.

Our analysis stresses the important role of the LHC/ILC interplay,
since neither of these colliders alone can provide us with the data needed
to determine the SUSY parameters in such scenarios with heavy sfermions
without tight model assumptions. \\[5mm]

\noindent{\bf Acknowledgments}\\[2mm]
We thank G.\ Polesello for many stimulating discussions.  JK and GMP
thank the Aspen Center for Physics where part of this work was
performed. JK and KR have been supported by the Polish Ministry of
Science and Higher Education Grant No~1~P03B~108~30 for the years 2006-2008 and
115/E-343/SPB/DESY/P-03/DWM517/2003-2005.

\end{document}